\documentclass[twocolumn,showpacs,preprintnumbers,amsmath,amssymb]{revtex4}
\usepackage{graphicx}
\usepackage{subfigure}
\usepackage{dcolumn}
\usepackage{bm}
\usepackage{mathrsfs}

\begin{document}

\title{The Second-Order Talbot Effect with Entangled Photon Pairs}

\date{\today}
\author{Kai-Hong Luo$^1$, Jianming Wen$^2$, Xi-Hao Chen$^1$, Qian Liu$^1$, Min Xiao$^2$ and Ling-An Wu$^1$}
\thanks{Corresponding author: wula@aphy.iphy.ac.cn}
\affiliation{$^1$Laboratory of Optical Physics, Institute of Physics
and Beijing National Laboratory for Condensed Matter Physics,
Chinese Academy of Sciences, Beijing 100190, China\\$^2$Department
of Physics, University of Arkansas, Fayetteville, Arkansas 72701,
USA}

\begin{abstract}
The second-order Talbot effect is analyzed for a periodic object
illuminated by entangled photon pairs in both the quantum imaging
and quantum lithography configurations. The Klyshko picture is
applied to describe the quantum imaging scheme, in which self-images
of the object that may or may not be magnified can be observed
nonlocally in the photon coincidences but not in the singles count
rate. In the quantum lithography setup, we find that the
second-order Talbot length is half that of the classical first-order
case, thus the resolution may be improved by a factor of two.
\end{abstract}


 \maketitle

\section{Introduction}

Classical image formation is most commonly associated with some type
of lens, with the exception of the earliest means of imaging known
to mankind, the pinhole camera. A less familiar form of lensless
imaging, in which a periodic structure can produce self-images at
certain regular distances, was first discovered by Talbot
\cite{Talbot} in 1836, partially explained by Rayleigh
\cite{Rayleigh} in 1881, and further explored by Weisel
\cite{Weisel} and Wolfke \cite{Wolfke} early in the last century.
The observation in recent years of this self-image replication in
several new areas of research has led to renewed interest in this
remarkable phenomenon, which contains much physics still unexplored.

The Talbot effect is a near field diffraction phenomenon in which a
plane wave transmitted through a grating or other periodic structure
propagates in such a way that the grating structure is replicated at
multiples of a certain longitudinal distance. Rayleigh explained it
as a natural consequence of Fresnel diffraction and the interference
of the diffracted beams  \cite{Rayleigh}; he showed analytically
that for plane wave illumination the self-images repeat at multiples
of the Talbot length  $z_{\textrm \tiny{T}}=2a^2/\lambda$, where $a$
is the period of the grating and $\lambda$ the wavelength of the
incident light. At half-integral multiples, the self-images are
laterally shifted by half a period, and at intermediate distances of
$z=(p/q)z_{\textrm \tiny{T}}$, where $p$ and $q$ are integers with
no common factor, the images have a smaller period of $a/n$
($n=2,3,4, \ldots$). These diffraction images are in fact a
superimposition of shifted and complex weighted replicas of the
original object \cite{Winthrop}.

Potential applications of the Talbot effect have been found in image
processing and synthesis, photolithography, optical testing, optical
metrology and spectrometry \cite{Patorski}. The effect is also well
known in the field of acousto-optics, electron optics and electron
microscopy \cite{Cowley}. Recently, it has also been demonstrated in
atomic waves \cite{Chapman,Wu}, Bose-Einstein condensates
\cite{Deng, Li}, large C$_{70}$ fullerene molecules \cite{Brezger},
waveguide arrays \cite{Iwanow}, and X-ray phase imaging
\cite{Weitkamp}.

However, all the above demonstrations involved the first-order field
intensity. In recent years, second-order (or two-photon) correlation
intensity measurements have led to the discovery of many new,
formerly unexpected phenomena. The first demonstration of ``ghost"
imaging and interference was performed by Shih's group in the mid
1990's, who exploited the spatial correlation of entangled photon
pairs generated from spontaneous parametric down conversion (SPDC)
\cite{Pittman,Strekalov}. In ghost imaging, an image of an object in
the signal path to one detector is observed in the coincidence
counts with another detector in the idler arm, if the Gaussian
``two-photon" thin lens equation is satisfied. A scheme with both
signal and idler beams in a Bessel beam mode was suggested by Vidal
et al. to demonstrate conditional interference patterns in the
second-order correlation function \cite{Vidal}. In this paper we
also exploit the transverse correlations of entangled photon pairs
from SPDC, and present a general analysis of the second-order Talbot
effect using the Klyshko advanced-wave picture \cite{Klyshko}. It
will be shown that self-images of a periodic structure can be
nonlocally observed in the two-photon coincidence count measurement
but not in the singles counts. Although spatially periodic objects
represent only a subgroup of all objects that can generate
self-images, they are still of fundamental importance and may reveal
interesting new physics.

\section{Two-photon Talbot Effect}

\subsection{Talbot effect in quantum imaging}

Let us first briefly review two-photon (biphoton) optics. From
Glauber's quantum measurement theory, the two-photon coincidence
counting rate for two point photodetectors is given by
\cite{Goodman}
\begin{eqnarray}
R_{cc}=\frac{1}{T}\int^T_0dt_1\int^T_0dt_2|\Psi(\vec{r}_1,t_1;\vec{r}_2,t_2)|^2,\label{eq:coincidence}
\end{eqnarray}
where the two-photon or biphoton amplitude $\Psi$ is determined by
the matrix element between the vacuum state $|0\rangle$ and the
two-photon state $|\psi\rangle$
\begin{eqnarray}
\Psi(\vec{r}_1,t_1;\vec{r}_2,t_2)=\langle0|E^{(+)}_1(\vec{r}_1,t_1)E^{(+)}_2(\vec{r}_2,t_2)|\psi\rangle.\label{eq:amplitude}
\end{eqnarray}
Here $E^{(+)}_j(\vec{r}_j,t_j)$ $(j=1,2)$  is the positive frequency
part of the electric field at point $\vec{r}_j$ on the $j$th
detector evaluated at time  $t_j$. We begin by computing the field
at the detector in terms of the photon destruction operators at the
output surface of the crystal:
\begin{eqnarray}
E^{(+)}_j(\vec{\rho}_j,z_j,t_j)&=&\int{d}\omega_j\int{d}^2\alpha_jE_jf_j(\omega_j)e^{-i\omega_jt_j}\nonumber\\
&\times&g_j(\vec{\alpha}_j,\omega_j;\vec{\rho}_j,z_j)a(\vec{\alpha}_j,\omega_j),\label{eq:field}
\end{eqnarray}
where $E_j=\sqrt{\hbar\omega_j/2\epsilon_0}$, $\vec{\alpha}_j$ is
the transverse wave vector, $f_j(\omega_j)$ is a narrow bandwidth
filter function peaked at central frequency $\Omega_j$, and the
Green's function $g_j(\vec{\alpha}_j,\omega_j;\vec{\rho}_j,z_j)$ is
the optical transfer function that describes the propagation of each
mode of angular frequency $\omega_j$ from the source to the
transverse point $\vec{\rho}_j$ in the plane of the $j$th detector
which is at a distance $z_j$ from the output surface of the crystal.
The photon annihilation operator $a(\vec{\alpha_j},\omega_j)$ at the
output surface of the source satisfies the commutation relation
\begin{eqnarray}
[a(\vec{\alpha},\omega),a^{\dag}(\vec{\alpha}',\omega')]=\delta(\vec{\alpha}-\vec{\alpha}')\delta(\omega-\omega').\label{eq:commutation}
\end{eqnarray}
We take the simplest model in which a plane-wave pump of frequency
$\omega_p$ and wave vector $k_p\hat{e}_z$ generates entangled
photons by SPDC in a crystal of length $L$. From perturbation
theory, the biphoton state at the output surface of the crystal
takes the form
\begin{eqnarray}
|\psi\rangle&=&\int{d}\omega_s\int{d}\omega_i\int{d}^2\alpha_1\int{d}^2\alpha_2\Phi(\omega_s,\omega_i)\nonumber\\&\times&\delta(\omega_s+\omega_i-\omega_p)\delta(\vec{\alpha}_1+\vec{\alpha}_2)|1_{\vec{k}_s},1_{\vec{k}_i}\rangle,\label{eq:state}
\end{eqnarray}
where $\Phi(\omega_s,\omega_i)$ is the spectral function resulting
from the phase matching, and $\omega_s$, $\omega_i$, $\vec{k}_s$,
and $\vec{k}_i$ are the frequencies and wave vectors of the
entangled signal and idler waves, respectively. The detailed form of
$\Phi$ is not important here since we are interested in the
transverse correlation of the photons. The $\delta$ functions
indicate that the source produces two-photon states with perfect
phase matching. We assume that the paraxial approximation holds and
the factors describing the temporal and transverse behavior of the
waves are separable. The frequency correlation determines the
two-photon temporal properties while the transverse momentum
correlation determines the spatial properties of the photon pairs.
It is the latter wave-vector correlation that is of prime interest
in second-order Talbot imaging.

We consider the setup shown in Fig.~1 to study the Talbot effect in
a typical quantum imaging configuration. The signal and idler
photons of wavelengths $\lambda_s$ and $\lambda_i$, respectively,
from SPDC in a nonlinear crystal are separated into two beams. An
object with a periodic structure is placed in the signal arm, and
the transmitted light collected by a bucket detector $D_s$. In the
idler arm the detector $D_i$, sometimes called the reference
detector, is a CCD camera or a scanning point detector, and
coincidence measurements are taken \cite{Klyshko}. The distance from
the output surface of the crystal to the object is $d_{s1}$, and to
the idler detector $d_i$; the distance between the object and
detector $D_s$ is $d_{s2}$.
\begin{figure}[tbp]
\includegraphics[scale=0.51]{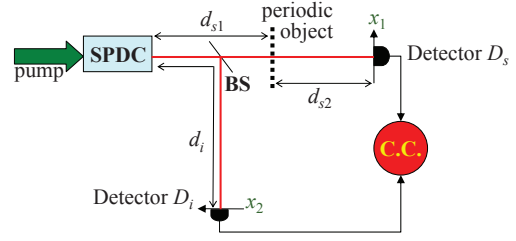}
\caption{(color online) Setup to show the Talbot effect in quantum
imaging using SPDC photons. BS, beamsplitter; C.C., coincidence
counter.}\label{fig1}
\end{figure}

Following the treatment in \cite{Goodman,Rubin,Wen}, we evaluate the
Green's functions $g_1(\vec{\alpha}_1,\omega_s;\vec{\rho}_1,d_s)$
and $g_2(\vec{\alpha}_2,\omega_i;\vec{\rho}_2,d_i)$ in the paraxial
approximation with an object described by the transparency function
$A_o(\vec{\rho}_o)$, and obtain
\begin{eqnarray}
g_1(\vec{\alpha}_1,\omega_s;\vec{\rho}_1,d_s)&=&-\frac{i\omega_s}{2\pi
cd_{s2}}e^{i\frac{\omega_s}{c}d_s}e^{i\frac{\omega_s}{2cd_{s2}}\rho^2_1}e^{-i\frac{cd_{s1}}{2\omega_s}\alpha^2_1}\nonumber\\&\times&\int{d}^2\rho_oA_o(\vec{\rho}_o)e^{i\frac{\omega_s\rho^2_o}{2cd_{s2}}}e^{i\vec{\rho}_o\cdot\left(\vec{\alpha}_1-\frac{\omega_s\vec{\rho}_1}{cd_{s2}}\right)},\label{eq:g1}\nonumber\\\
g_2(\vec{\alpha}_2,\omega_i;\vec{\rho}_2,d_i)&=&e^{i\frac{\omega_i}{c}d_i}e^{-i\frac{cd_i}{2\omega_i}\alpha^2_2}e^{i\vec{\alpha}_2\cdot\vec{\rho}_2},\label{eq:g2}
\end{eqnarray}
where $d_s=d_{s1}+d_{s2}$ and $A_o(\vec{\rho}_o)$ is the aperture
function of the object. From Eqs.~(\ref{eq:field}) and
(\ref{eq:state}), and with the assumption that
$\omega_j=\Omega_j+\nu_j$, where $|\nu_j|\ll\Omega_j$ and
$\Omega_s+\Omega_i=\omega_p$, the temporal (longitudinal) and
transverse terms can be factored out so the two-photon amplitude in
Eq.~(\ref{eq:amplitude}) becomes
\begin{eqnarray}
\Psi(\vec{r}_1,t_1;\vec{r}_2,t_2)=e^{i(\Omega_s\tau_1+\Omega_i\tau_2)}\psi(\vec{\rho}_1,\tau_1;\vec{\rho}_2,\tau_2),\label{eq:amplitude2}
\end{eqnarray}
where $\tau_1=t_1-d_s/c$, $\tau_2=t_2-d_i/c$, and
\begin{eqnarray}
\psi(\vec{\rho}_1,\tau_1;\vec{\rho}_2,\tau_2)&=&\int{d}\nu_s\int{d}\nu_i\delta(\nu_s+\nu_i)e^{i(\nu_s\tau_1+\nu_i\tau_2)}\nonumber\\&\times&
f_1(\Omega_s+\nu_s)f_2(\Omega_i+\nu_i)B(\vec{\rho}_1,\vec{\rho}_2).\label{eq:decomposition}
\end{eqnarray}
The transverse part $B(\vec{\rho}_1,\vec{\rho}_2)$ takes the form
\begin{eqnarray}
B(\vec{\rho}_1,\vec{\rho}_2)&=&B_0\int{d}^2\rho_oA_o(\vec{\rho}_o)e^{i\frac{\omega_s}{2cd_{s2}}\rho^2_o}e^{-i\frac{\omega_s}{cd_{s2}}\vec{\rho}_o\cdot\vec{\rho}_1}\nonumber\\&\times&
\int{d}^2\alpha_1e^{-i\frac{c}{2}\alpha^2_1\left(\frac{d_{s1}}{\omega_s}+\frac{d_i}{\omega_i}\right)}e^{i\vec{\alpha}_1\cdot(\vec{\rho}_o-\vec{\rho}_2)},\label{eq:transverse}
\end{eqnarray}
where we have collected all the slowly varying terms into the
constant $B_0$. Completing the integration on the transverse mode
$\vec{\alpha}_1$ in Eq.~(\ref{eq:transverse}) gives
\begin{eqnarray}
B(\vec{\rho}_1,\vec{\rho}_2)&=&B_0\int{d}^2\rho_oA_o(\vec{\rho}_o)\nonumber\\&\times&\textrm{exp}\left(-i\frac{\omega_s}{c}\vec{\rho}_o\cdot\left(\frac{\vec{\rho}_1}{d_{s2}}+\frac{\vec{\rho}_2}{d_{s1}+\frac{\omega_s}{\omega_i}d_i}\right)\right)\nonumber\\&\times&\textrm{exp}\left(i\frac{\omega_s}{2c}\rho^2_o\left(\frac{1}{d_{s2}}+\frac{1}{d_{s1}+\frac{\omega_s}{\omega_i}d_i}\right)\right).\label{eq:transverse2}
\end{eqnarray}
Here again all the irrelevant constants have been absorbed into
$B_0$.

For simplicity, our discussion will be restricted to one-dimensional
objects, but extension of the analysis to two-dimensional objects is
straightforward. The transmission function for a general
one-dimensional periodic object can be expanded as a Fourier series
\begin{eqnarray}
A_o(x)=\sum^{\infty}_{n=-\infty}c_ne^{-i\frac{2\pi{n}x}{a}},\label{eq:object}
\end{eqnarray}
where $a$ is the spatial period along the transverse $x$ direction
and $c_n$ is the coefficient of the $n$th harmonic. We shall not
specify the form of $c_n$ at this point, so any type of periodic
object can be assumed for the present analysis. By substituting
Eq.~(\ref{eq:object}) into Eq.~(\ref{eq:transverse2}) and working in
Cartesian coordinates, the transverse part of the two-photon
amplitude in Eq.~(\ref{eq:transverse2}) can be written as
\begin{eqnarray}
B(x_1,x_2)&=&B_0\sum^{\infty}_{n=-\infty}c_n\textrm{exp}\left(-i\frac{\frac{n^2\pi}{a^2}\lambda_s}{\frac{1}{d_{s2}}+\frac{1}{d_{s1}+\frac{\lambda_i}{\lambda_s}d_i}}\right)\nonumber\\
&\times&\textrm{exp}\left(i\frac{\frac{2\pi
n}{a}\left(\frac{x_1}{d_{s2}}+\frac{x_2}{d_{s1}+\frac{\lambda_i}{\lambda_s}d_i}\right)}{\frac{1}{d_{s2}}+\frac{1}{d_{s1}+\frac{\lambda_i}{\lambda_s}d_i}}\right).\label{eq:transverse4}
\end{eqnarray}
The exponential term in Eq.~(\ref{eq:transverse4}) is of basic
importance to self-imaging, and will be called the ``localization"
term, since it describes the phase changes of the diffraction orders
along the directions of propagation. It is known that self-imaging
occurs in planes where the transmitted object light amplitudes are
repeated, that is, when all diffraction orders are in phase and
interfere constructively. From Eq.~(\ref{eq:transverse4}) we see
that this can occur at certain distances when the first term equals
$1$ for all $n$, that is,
\begin{eqnarray}
\frac{1}{\lambda_sd_{s2}}+\frac{1}{\lambda_sd_{s1}+\lambda_id_i}=\frac{1}{2ma^2},\label{eq:talbotlength1}
\end{eqnarray}
where $m$ is an integer referred to as the self-imaging number.
Defining $z_{s\textrm \tiny{T}}=2a^2/\lambda_s$, we can rewrite
Eq.~(\ref{eq:talbotlength1}) as
\begin{eqnarray}
\frac{1}{d_{s2}}+\frac{1}{d_{s1}+\frac{\lambda_i}{\lambda_s}d_i}=\frac{1}{mz_{s\textrm
\tiny{T}}},\label{eq:talbotlength2}
\end{eqnarray}
Comparing this with the well-known Gaussian thin lens equation, we
can consider the self-imaging to be a counterpart phenomenon in
which a sequence of lenses of focal length $z_{s\textrm \tiny{T}}$
is situated in the plane of the periodic object and produces images
of the source. The distance $z_{s\textrm \tiny{T}}$ may be regarded
as the Talbot length for second-order correlation quantum imaging.

We note that the factor $2$ in Eq.~(\ref{eq:talbotlength1}) is
inconsequential, and so may be omitted. Thus in the cases when $m$
is an odd integer, there will be self-images with a lateral shift of
half a period relative to the object, due to the $\pi$ phase shift
of odd-number diffraction orders relative to the zero and
even-number orders.

As mentioned in the previous section, in the self-image planes
Eq.~(\ref{eq:transverse4}) reduces to
\begin{eqnarray}
B(x_1,x_2)=B_0\sum^{\infty}_{n=-\infty}c_ne^{\left[i\frac{2\pi{n}}{a}\frac{\left(d_{s1}+\frac{\lambda_i}{\lambda_s}d_i\right)x_1+d_{s2}x_2}{d_{s1}+d_{s2}+\frac{\lambda_i}{\lambda_s}d_i}\right]},\label{eq:talbot1}
\end{eqnarray}
which carries information about the lateral magnification $M$ of the
imaging patterns arising as a result of the nonlocal correlation
between entangled photons.

In a typical quantum imaging experiment, if the positions of the
object and lens are specified, the lens equation immediately tells
us the relationships between the image, object, and lens. This leads
to the intuitive Klyshko interpretation of two-photon geometrical
optics \cite{Klyshko}, in which one of the photons is created at the
detector placed in the same arm as the object, propagates back to
the source where it becomes the second photon, which then propagates
forward in time to the other detector. However, this picture may not
be directly applicable to the second-order Talbot effect, because
here, in principle, both detectors can be regarded as the source,
and there are many complex cases with different possible
magnifications. Nevertheless, there are two simple cases in which
the Klyshko picture may be directly used to interpret the imaging
process, that is, when either one of the detectors is fixed at its
origin. This corresponds to either $x_1=0$ or $x_2=0$ in
Eq.~(\ref{eq:talbot1}), in which case the images will be magnified
by a factor of $1+\frac{d_{s1}+\lambda_i/\lambda_sd_i}{d_{s2}}$ or
$1+\frac{d_{s2}}{d_{s1}+\lambda_i/\lambda_sd_i}$, respectively.
Indeed, since both detectors could be visualized as point sources,
two sets of self-images would be observable in coincidence
measurements. The magnification for the images in such a case is not
easy to describe since it depends on the detection scheme.

Before proceeding to the next section, we should check whether the
Talbot effect could be observed with a single detector, so we should
analyse the photon statistics in detector $D_s$ of Fig.~1. The
single-photon count rate is defined as
\begin{eqnarray}
R_s=\frac{1}{T}\int^T_0dt_1\sum_{\vec{k}_i}|\langle0|a(\vec{k}_i)E^{(+)}_1(\vec{r}_1,t_1)|\psi\rangle|^2.\label{eq:singlerate}
\end{eqnarray}
Assuming there are infinite transverse modes in the SPDC process,
with the use of Eqs.~(\ref{eq:field}), (\ref{eq:state}), and
(\ref{eq:g1}) we find that in the transverse plane the single-photon
counts is proportional to $ \int{d}^2\rho_o|A_o(\vec{\rho}_o)|^2$,
so there is no diffraction image if only one detector is used.

\subsection{Talbot effect in quantum lithography}

In the proof-of-principle quantum lithography experiment of
\cite{Boto}, a factor of two over the classical diffraction limit
was demonstrated by utilizing the entangled nature of a two-photon
state. In this section we will extend our analysis to a quantum
lithography experimental setup and see whether there is any
difference in the Talbot self-imaging effect. In the scheme of
Fig.~2, the periodic object is placed immediately behind the SPDC
source, and we assume that the degenerate signal and idler photons
($\lambda_s=\lambda_i$) are detected by two single-photon detectors
$D_s$ and $D_i$, or a two-photon detector without the beamsplitter.
The distance between the object and the detectors is $d_0$.
Following \cite{Goodman,Rubin,Wen}, it is straightforward to show
that the Green's function for the signal ($j=1$) and idler ($j=2$)
photons now takes the form of
\begin{eqnarray}
g_j(\vec{\alpha}_j,\omega;\vec{\rho}_j,d)&=&-i\frac{e^{i\frac{\omega{d}}{c}}}{\lambda{d}}e^{i\frac{\omega\rho_j^2}{2cd}}\int{d}^2\rho_oA_o(\vec{\rho}_o)e^{i\frac{\omega\rho^2_o}{2cd}}\nonumber\\&\times&
e^{i\vec{\rho}_o\cdot(\vec{\alpha}_j-\frac{\omega\vec{\rho}_j}{cd})}.\label{eq:greenfunction}
\end{eqnarray}
By applying the same procedure in Sec.~II A, the transverse part of
the two-photon amplitude (\ref{eq:decomposition}) now becomes
\begin{eqnarray}
B(\vec{\rho}_1,\vec{\rho}_2)=B_0\int{d}^2\rho_oA^2_o(\vec{\rho}_o)e^{i\frac{\omega\rho^2_o}{cd}}e^{-i\frac{\omega\vec{\rho}_o\cdot(\vec{\rho}_1+\vec{\rho}_2)}{cd}}.\label{eq:lithography1}
\end{eqnarray}

\begin{figure}[tbp]
\includegraphics[scale=0.51]{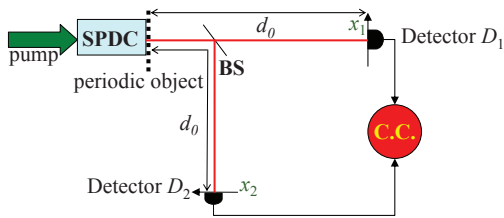}
\caption{(color online) Scheme to demonstrate the Talbot effect
using degenerate SPDC photons in the quantum lithography
configuration. BS, beamsplitter; C.C., coincidence
counter.}\label{fig2}
\end{figure}

For simplicity, we still consider the one-dimensional periodic
object given in Eq.~(\ref{eq:object}). Plugging
Eq.~(\ref{eq:object}) into (\ref{eq:lithography1}), we obtain
\begin{eqnarray}
B(x_1,x_2)=B_0\sum^{\infty}_{n=-\infty}c^2_ne^{-i2\pi\frac{{n}^2}{a^2}\lambda
d_0}e^{i\frac{2\pi{n}}{a}(x_1+x_2)}.\label{eq:lithography3}
\end{eqnarray}
As in our previous discussion, the condition for revival patterns of
the periodic structure to occur is that the diffraction orders have
equal phases when the distances $d_0$ satisfies
\begin{eqnarray}
d_0=\frac{ma^2}{\lambda}=mz_{s\textrm
\tiny{T}},\label{eq:lithotalbotlength1}
\end{eqnarray}
and Eq.~(\ref{eq:lithography3}) then reduces to
\begin{eqnarray}
B(x_1,x_2)=B_0\sum^{\infty}_{n=-\infty}c^2_ne^{i\frac{2\pi{n}}{a}(x_1+x_2)}.\label{eq:lithography4}
\end{eqnarray}
This intensity-intensity correlation distribution is almost the same
as that at the exit surface of the object. Similarly, as in the
previous case, the self-imaging planes can be obtained by setting
either detector at the origin. If the factor $m$ in
Eq.~(\ref{eq:lithotalbotlength1}) is an odd integer, the
self-replicated diffraction pattern is also laterally displaced by
half a period with respect to the original object.

We notice a number of interesting points from
Eqs.~(\ref{eq:lithotalbotlength1}) and (\ref{eq:lithography4}).
First, for a source of the same wavelength, $z_{s\textrm \tiny{T}}$
is only half of the classical first-order Talbot length $z_{\textrm
\tiny{T}}$, but there is no magnification in the image. Secondly,
for collinear degenerate SPDC as considered here, both entangled
photons pass through the same spatial point in the object, so the
Fourier coefficients $c_n$ become $c^2_n$. Thirdly, if a detector
capable of resolving two photons were employed, an enhancement
factor of $2$ in the self-images would be obtained compared with the
classical first-order case.

\subsection{ Numerical Examples and Discussion}

\begin{figure}[tbp]
\includegraphics[scale=0.41]{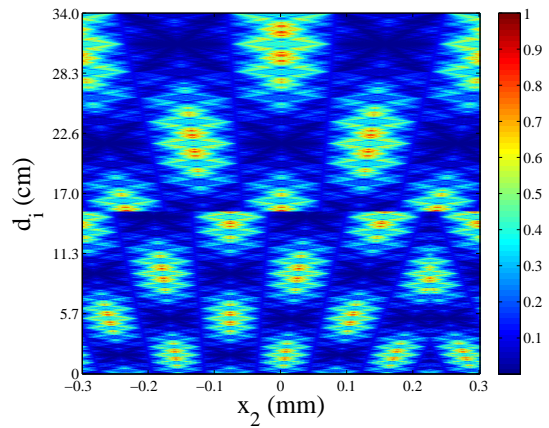}
\caption{(color online) Second-order Talbot imaging carpet obtained
by scanning the idler detector $D_i$ through $d_i=$ 0 -- 34 cm along
the longitudinal $z$ direction and through $x_2=-$ 0.3 -- 0.3 mm in
the transverse $x$ direction while keeping the signal detector $D_s$
fixed at position $d_{s2}=20$ cm and $x_1=0$, and $d_{s1}=11$ cm.
The color bar denotes the transverse value of the two-photon
correlation function.} \label{fig3}
\end{figure}

\begin{figure}
\centering \subfigure [~Second-order Talbot imaging carpet]{
\label{fig4:subfig:a} 
\includegraphics[scale=0.41]{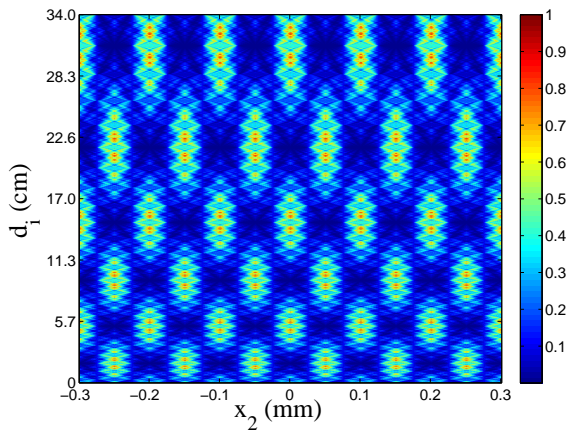}}
\hspace{1in} \subfigure [~Second-order Talbot image]{
\label{fig4:subfig:b} 
\includegraphics[scale=0.41]{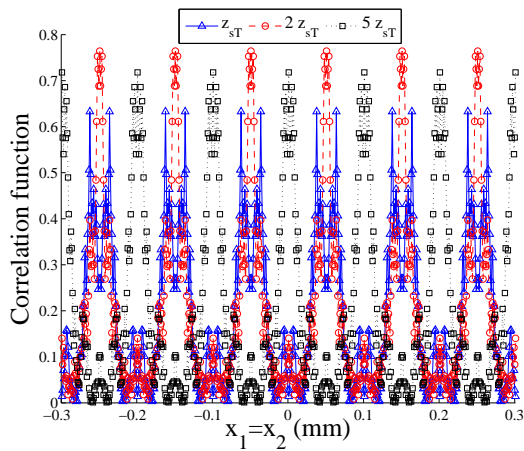}}
\caption{(color online) (a) Second-order Talbot imaging carpet
obtained by moving both detectors synchronously along the transverse
$x$ direction ($x_1=x_2=-$ 0.3 -- 0.3 mm) and scanning the idler
detector $D_i$ through $d_i=$ 0 -- 34 cm along the $z$ direction
while keeping the signal detector $D_s$ fixed at position
$d_{s2}=20$ cm, with $d_{s1}=11$ cm. The color bar denotes the value
of the transverse two-photon correlation function. (b) Second-order
Talbot image obtained as in (a), but with $D_i$ at positions
$d_i=z_{s\textrm \tiny{T}}$ (blue triangles), $d_i=2 z_{s\textrm
\tiny{T}}$ (red circles) and $d_i=5 z_{s\textrm \tiny{T}}$ (black
squares).}
\end{figure}

For convenience, we assume that both idler and signal photons are
generated by SPDC at the wavelength
$\lambda_s=\lambda_i=\lambda=883.2$ nm when light from a $441.6$ nm
pump laser beam is incident on a nonlinear crystal, and the periodic
object is a one dimensional rectangular grating $a=0.1$ mm. In the
quantum imaging scheme, the main result of the second-order Talbot
effect is represented by Eq.~(\ref{eq:transverse4}). Figure~3 shows
the numerically computed second-order correlation pattern obtained
by scanning the idler detector $D_i$ along the longitudinal $z$ and
transverse $x$ directions while keeping the signal detector ($D_s$)
fixed at position $d_{s2}$ and $x_1=0$, the distance $d_{s1}$ from
the crystal to the periodic object also being fixed. We see that a
typical Talbot ``carpet" pattern is produced. In the same way,
Fig.~4 (a) shows the second-order Talbot imaging carpet obtained
when both detectors are moved synchronously along the transverse $x$
plane, with the signal detector $D_s$ fixed at position $d_{s2}=20$
cm, and the idler detector $D_i$ scanned along the longitudinal $z$
direction. The transverse profile of the ``carpet" is given by the
second-order Talbot image of Fig.~4 (b),where the self-image is
obtained by  moving both detectors synchronously in the $x$
direction, keeping $d_{s2}$ fixed, and the longitudinal positions of
$D_i$ at $d_i=z_{s\textrm \tiny{T}}$ (blue triangles), $d_i=2
z_{s\textrm \tiny{T}}$ (red circles), and $d_i=5 z_{s\textrm
\tiny{T}}$ (black squares). It is found that the second-order Talbot
image produced by scanning only one detector is magnified, while
that obtained by moving both detectors synchronously is the same
size as the original object. It is interesting to note that the
latter case holds even if the detectors are not at the same distance
from the light source.

There are also some interesting features that may be observed from
Eq.~(\ref{eq:lithography3}) in the quantum lithography case. The
second-order Talbot length $z_{s\textrm \tiny{T}}=a^2/\lambda$ is
only half the classical Talbot length $z_{\textrm
\tiny{T}}=2a^2/\lambda$. Figure 5 shows the second-order
interference pattern obtained by scanning both detectors along the
$z$ direction in the region of $d_0=$ 0 -- 5.7 cm while keeping the
transverse position of one detector ($D_s$ or $D_i$) fixed at
$x_1=0$ (or $ x_2=0$), and scanning the other detector in the
transverse direction. When both detectors are moved together along
the transverse direction, we find that the period of the grating is
halved, that is, the longitudinal resolution is enhanced by a factor
of two. Moreover, when both detectors are displaced simultaneously
in the same manner along the $x$ and $z$ directions, then not only
is the longitudinal resolution increased but also the spatial
resolution is improved by a factor of two, as shown in Fig.~6.

\begin{figure}[tbp]
\includegraphics[scale=0.41]{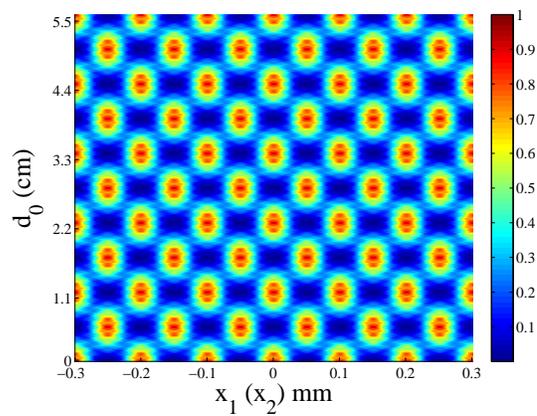}
\caption{(color online) Second-order Talbot lithography carpet
obtained by scanning both detectors $D_i$ and $D_s$ along the $z$
direction from $d_0=$ 0 -- 5.7 cm with the transverse position of
one detector fixed at $x_1(x_2)=0$, and the other detector scanned
in the transverse $x_2(x_1)$ direction. The color bar denotes the
value of the transverse two-photon correlation
function.}\label{fig5}
\end{figure}

\begin{figure}[tbp]
\includegraphics[scale=0.41]{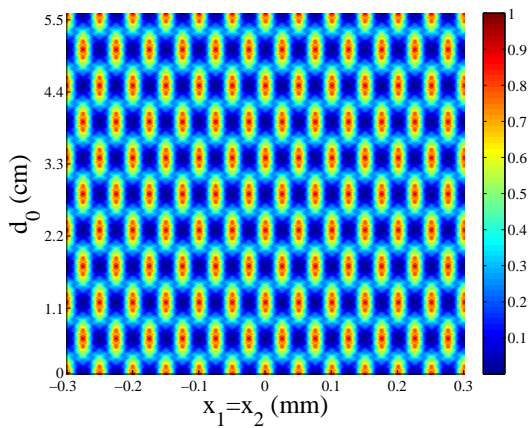}
\caption{(color online) Second-order Talbot lithography carpet
obtained by moving both detectors simultaneously in the same manner
along the transverse $x$ and longitudinal $z$ directions. The
transverse scanning range is $x_1=x_2=-$ 0.3 -- 0.3 mm. The color
bar denotes the value of the transverse two-photon correlation
function.}\label{fig6}
\end{figure}
In the above derivations we have implicitly assumed normal incidence
of the illuminating fields to the object plane. It is possible to
generalize our analysis to the case of non-normal incidence. A
lateral displacement of the SPDC source from the optical axis would
introduce a displacement of the observed second-order self-images
proportional to the observation distance and incident angle
\cite{Patorski}. The same would be true for a nonparallel
illumination beam. Moreover, the analysis throughout the paper is
based on a plane-wave input beam, but it is possible to take into
account the Gaussian profile of the pump.

It should be emphasized that the results obtained in Sections II A
and B are valid when the parabolic approximation is satisfied for
all diffraction orders. This implies that all diffraction orders are
simultaneously co-phasial in the self-imaging planes. In classical
wave optics, this problem was addressed, for example, by
Sciammarella and Davis \cite{Scimmarella} and Chang \cite{Chang}. By
setting a tolerance on the maximum allowable optical path difference
introduced by the third term in the binomial expansion of
$\sqrt{1+(\rho/z)^2}$, one can determine the maximum number $m$ of
the self-images formed by a specified diffraction order $n$ of the
object. At larger observation distances, due to gradual violation of
the paraxial approximation by higher diffraction orders, one can no
longer speak of ``self-imaging" in the strict sense. Experimentally,
the images would become more and more blurred.

Before concluding this section, we should remark that the Talbot
self-imaging effect with entangled photon pairs studied here can be
generalized to multiphoton entangled states \cite{Wen,Wen2}.
Although the realization of entangled multiphoton sources is
experimentally challenging, the inherent physics would be of great
richness and well worth studying.

\section{Summary}

In summary, we have theoretically studied quantum ghost imaging with
a periodic structure as the object, illuminated by entangled photon
pairs generated from SPDC with a plane-wave incident pump. It is
shown that a second-order Talbot effect (self-imaging effect) in
both the quantum imaging and lithography configurations may be
observed, without the need of any focussing lens. The self-images
observed in the two-photon coincidence counts are nonlocal, and
there is no imaging in the single-photon counts. In the quantum
imaging setup, the Klyshko picture of two-photon geometric optics
may be applied to interpret the imaging process under certain
conditions. In the quantum lithographic configuration, we show that
the Talbot length is half of that in the classical first-order case.

We also notice that the quadratic phases shown in the origin of the
Talbot effect have played an important role in revivals and
fractional revivals \cite{Leichtle}, curlicues \cite{Berry}, quantum
carpets \cite{Marzoli}, and Gaussian sums \cite{Mack}. It is to be
hoped that many other interesting second-order phenomena based on
entangled photons may result from our above analysis.

\section{Acknowledgments}
We thank Sai-Jun Wu and Morton H. Rubin for helpful discussions.
This work was supported by the National Natural Science Foundation
of China (Grant No 10674174) and the National Program for Basic
Research in China (Grant 2006CB921107). J.W. and M.X. acknowledge
partial support from the National Science Foundation of the U.S.A.


\begin{thebibliography}{99}
\bibitem{Talbot} H. F. Talbot, Philos. Mag. \textbf{9}, 401 (1836).

\bibitem{Rayleigh} L. Rayleigh, Philos. Mag. \textbf{11}, 196 (1881).

\bibitem{Weisel} H. Weisel, Ann. Phys. Lpz. \textbf{33}, 995 (1910).

\bibitem{Wolfke} M. Wolfke, Ann. Phys. Lpz. \textbf{40}, 194 (1913).

\bibitem{Winthrop} J. T. Winthrop and C. R. Worthington, J. Opt. Soc. Am. \textbf{55}, 373-381 (1965)

\bibitem{Patorski} K. Patorski, in \textit{Progress in Optics} \textbf{27}, pp. 1-108, edited by E. Wolf (North-Holland, Amsterdam, 1989).

\bibitem{Cowley} J. M. Cowley and A. F. Moodie, Proc. Phys. Soc. B \textbf{70}, 486 (1957); \textbf{70}, 497 (1957); \textbf{70}, 505 (1957); J. M. Cowley, \textit{Diffraction Physics} (North-Holland, Amsterdam, 1995).

\bibitem{Chapman} M. S. Chapman, C. R. Ekstrom, T. D. Hammond, J. Schmiedmayer, B. E. Tannian, S. Wehinger, and D. E. Pritchard, Phys. Rev. A \textbf{51}, R14 (1995).

\bibitem{Wu} S. Wu, E. Su, and M. Prentiss, Phys. Rev. Lett. \textbf{99}, 173201 (2007).

\bibitem{Deng} L. Deng, E. W. Hagley, J. Denschlag, J. E. Simsarian, M. Edwards, C. W. Clark, K. Helmerson, S. L. Rolston, and W. D. Phillips, Phys. Rev. Lett. \textbf{83}, 5407 (1999).

\bibitem{Li} Ke Li, L. Deng, E.W. Hagley, M. G. Payne, and M. S. Zhan, Phys. Rev. Lett. \textbf{101}, 250401 (2008).

\bibitem{Brezger} B. Brezger, L. Hackem\"{u}ller, S. Uttenthaler, J. Petschinka, M. Arndt, and A. Zeilinger, Phys. Rev. Lett. \textbf{88}, 100404 (2002).

\bibitem{Iwanow} R. Iwanow, D. A. May-Arrioja, D. N. Christodoulides, G. I. Stegeman, Y. Min, and W. Sohler, Phys. Rev. Lett. \textbf{95}, 053902 (2005).

\bibitem{Weitkamp} T. Weitkamp, B. N\"{o}hammer, A. Diaz, and C. David, Appl. Phys. Lett. \textbf{86}, 54101-54103 (2005).

\bibitem{Pittman} T. B. Pittman, Y.-H. Shih, D. V. Strekalov, and A. V. Sergienko, Phys. Rev. A \textbf{52}, R3429 (1995).

\bibitem{Strekalov} D. V. Strekalov, A. V. Sergienko, D. N. Klyshko, and Y.-H. Shih, Phys. Rev. Lett. \textbf{74}, 3600 (1995).

\bibitem{Vidal} I. Vidal, S. B. Cavalcanti, E. J. S. Fonseca, and J. M. Hickmann, Phys. Rev. A \textbf{78}, 033829 (2008).

\bibitem{Klyshko} D. N. Klyshko, Phys. Lett. A \textbf{128}, 1337 (1988); Sov. Phys. Usp. \textbf{31}, 74 (1988).

\bibitem{Goodman} J. W. Goodman, \textit{Introduction to Fourier Optics} (McGraw-Hill, New York, 1968).

\bibitem{Rubin} M. H. Rubin, Phys. Rev. A \textbf{54}, 5349 (1996).

\bibitem{Wen} J.-M. Wen, P. Xu, M. H. Rubin, and Y.-H. Shih, Phys. Rev. A \textbf{76}, 023828 (2007).

\bibitem{Boto} A. N. Boto, P. Kok, D. S. Abrams, S. L. Braunstein, C. P. Williams,
and J. P. Dowling, Phys. Rev. Lett. \textbf{85}, 2733 (2000)

\bibitem{Scimmarella} C. A. Scimmarella and D. Davis, Exp. Mech. \textbf{8}, 459 (1968).

\bibitem{Chang} B. J. Chang, Ph.D. Thesis, University of Michigan (1974).

\bibitem{Wen2}J.-M. Wen and M. H. Rubin, Phys. Rev. A 79, 025802 (2009).

\bibitem{Leichtle} C. Leichtle, I. S. Averbukh, and W. P. Schleich, Phys. Rev. Lett. \textbf{77}, 3999 (1996).

\bibitem{Berry} M. V. Berry and J. Goldberg, Nonlinearity \textbf{1}, 1 (1988); M. V. Berry and S. Klein, J. Mod. Opt. \textbf{43}, 2139 (1996).

\bibitem{Marzoli} M. V. Berry, I. Marzoli, and W. P. Schleich, Phys. World \textbf{14}, 39 (2001); O. Friesch, I. Marzoli, and W. P. Schleich, New J. Phys. \textbf{2}, 4 (2000).

\bibitem{Mack} H. Mack, M. Bienert, F. Haug, M. Freyberger, and W. P. Schleich, Phys. Stat. Sol. (b) \textbf{233}, 408 (2002).

\end{thebibliography}
\end{document}